\documentclass[sigconf]{acmart}

\usepackage{balance}
\usepackage{amsmath}
\usepackage{url}

\usepackage{algorithm}
\usepackage{algpseudocode}
\usepackage{textcomp}
\usepackage{xcolor}

\usepackage{graphicx}
\usepackage{mathtools}
\usepackage{tikz}
\usepackage[export]{adjustbox}
\usepackage{caption}
\usepackage{multirow}
\usepackage{ragged2e}

\usepackage{subcaption}
\usepackage[normalem]{ulem}
\useunder{\uline}{\ul}{}
\usepackage[inline]{enumitem}
\usepackage[most]{tcolorbox}

\AtBeginDocument{%
  }

\copyrightyear{2025} 
\acmYear{2025} 
\setcopyright{cc}
\setcctype{by}
\acmConference[CIKM '25]{Proceedings of the 34th ACM International
Conference on Information and Knowledge Management}{November 10--14,
2025}{Seoul, Republic of Korea}
\acmBooktitle{Proceedings of the 34th ACM International Conference on
Information and Knowledge Management (CIKM '25), November 10--14, 2025,
Seoul, Republic of Korea}\acmDOI{10.1145/3746252.3760878}
\acmISBN{979-8-4007-2040-6/2025/11}

\begin{document}

\title{Temporal-Aware User Behaviour Simulation with Large Language Models for Recommender Systems}

\author{Xinye Wanyan}
\email{xinye.wanyan@student.rmit.edu.au}
\orcid{0009-0002-7264-1803}
\affiliation{%
  \institution{RMIT University}
  \city{Melbourne}
  \state{VIC}
  \country{Australia}
}
\author{Danula Hettiachchi}
\email{danula.hettiachchi@rmit.edu.au}
\orcid{0000-0003-3875-5727}
\affiliation{%
  \institution{RMIT University}
  \city{Melbourne}
  \state{VIC}
  \country{Australia}
}
\author{Chenglong Ma}
\email{chenglong.ma@rmit.edu.au}
\orcid{0000-0002-6745-4029}
\affiliation{%
  \institution{RMIT University}
  \city{Melbourne}
  \state{VIC}
  \country{Australia}
}
\author{Ziqi Xu}
\email{ziqi.xu@rmit.edu.au}
\orcid{0000-0003-1748-5801}
\affiliation{%
  \institution{RMIT University}
  \city{Melbourne}
  \state{VIC}
  \country{Australia}
}

\author{Jeffrey Chan}
\email{jeffrey.chan@rmit.edu.au}
\orcid{0000-0002-7865-072X}
\affiliation{%
  \institution{RMIT University}
  \city{Melbourne}
  \state{VIC}
  \country{Australia}
}

\begin{abstract}
    Large Language Models (LLMs) demonstrate human-like capabilities in language understanding, reasoning, and generation, driving interest in using LLM-based agents to simulate human feedback in recommender systems. However, most existing approaches rely on static user profiling, neglecting the temporal and dynamic nature of user interests. This limitation stems from a disconnect between language modelling and behaviour modelling, which constrains the capacity of agents to represent sequential patterns. To address this challenge, we propose a \underline{Dy}namic \underline{T}emporal-aware \underline{A}gent-based simulator for \underline{Rec}ommender Systems, DyTA4Rec, which enables agents to model and utilise evolving user behaviour based on historical interactions. 
    DyTA4Rec features a dynamic updater for real-time profile refinement, temporal-enhanced prompting for sequential context, and self-adaptive aggregation for coherent feedback.
    % DyTA4Rec includes a dynamic updater for refining user profiles in real time, a temporal-enhanced prompting strategy for incorporating sequential context, and a self-adaptive aggregation mechanism for producing coherent feedback.
    Experimental results at group and individual levels show that DyTA4Rec significantly improves the alignment between simulated and actual user behaviour by modelling dynamic characteristics and enhancing temporal awareness in LLM-based agents.\footnote{~The source code can be found at~\url{https://github.com/WennyXY/DyTA4Rec}.}

\end{abstract}

\begin{CCSXML}
<ccs2012>
   <concept>
       <concept_id>10002951.10003317.10003347.10003350</concept_id>
       <concept_desc>Information systems~Recommender systems</concept_desc>
       <concept_significance>500</concept_significance>
       </concept>
 </ccs2012>
\end{CCSXML}

\ccsdesc[500]{Information systems~Recommender systems}

\keywords{Recommender System, Large Language Model, Generative Agents}

\maketitle

\section{Introduction}
% % Recommendation system has emerged as a leading method for the automated culture curating and determining the information to which users are exposed, thus significantly shaping their minds \cite{adomavicius2019hidden}.
% Modern recommender systems demand real-time interactions to enable effective performance assessment and iterative system improvement \cite{gao2022kuairec}.
% % Traditional simulators \cite{shi2018virtualtaobao, zhao2023kuaisim} constrained by predefined behavior templates based on static historical interactions can provide consistent and believable reactions.
% % In practice, it turns out to be resources-intensive and not enough for the vast and complex human behaviour space.
% Recently, large language models (LLMs) with demonstrated human-level intelligence in general-purpose understanding, reasoning, and decision-making \cite{zhao2023survey,radford2019language} have been employed as credible human proxies in interactions with recommender systems \cite{zhang2024agentcf,zhang2024generative,wang2025user}. 
% LLM-based agents enable dynamic and nuanced user profiling and facilitate real-time and diverse evaluation with minimum task-specific training.
% Specifically, LLMs are utilised as the central reasoning component equipped with modules for user profiling, memory management, and action selection, to support behaviour simulation in recommender systems.

Modern recommender systems (RSs) demand real-time interactions to enable effective evaluation and iterative system improvement~\cite{gao2022kuairec,ZhangYCLLXZ25,xu2025towards}. To meet this need, recent studies have explored the use of large language models (LLMs), which exhibit human-level capabilities in general-purpose understanding, reasoning, and decision making~\cite{zhao2023survey,radford2019language}, as credible proxies for interacting with RSs~\cite{zhang2024agentcf,zhang2024generative,wang2025user,zhang2025llm,zhu2025llm,chen2025recusersim,ma2025pub}. LLM-based agents support dynamic and nuanced user profiling and enable real-time, diverse evaluation with minimal task-specific training. Specifically, LLMs function as the central reasoning module, equipped with components for user profiling, memory management, and action selection, to simulate user behaviour in RSs \cite{zhang2025survey}.

Building upon this general LLM-based agent framework, various agent architectures are developed and customised for specific application scenarios to enhance the fidelity of user behaviour simulation in RSs. Agent4Rec~\cite{zhang2024generative} performs statistical analysis on user interaction patterns and incorporates the extracted insights into user profiles, thereby improving the realism of simulated behaviours. Its action module also accounts for both user preferences and emotion-driven responses. Extending beyond individual user modelling, AgentCF~\cite{zhang2024agentcf} introduces user and item agents that engage in interactive processes, enabling them to collaboratively learn and adapt through mutual interaction. In contrast, RecAgent~\cite{wang2025user} focuses on modelling the influence of social context by integrating external relationships and social activities into the simulation of user responses to recommendations.

Existing frameworks primarily focus on inferring user preferences and characteristics from static historical data, overlooking the dynamic evolution of user interests and the temporal patterns underlying user behaviour. Prior research has shown that incorporating temporal dynamics into user modelling can significantly improve the accuracy and effectiveness of personalisation strategies~\cite{bogina2023considering}. However, recent work by~\citet{hou2024large} shows that LLMs have notable limitations in modelling sequential patterns, which can hinder their ability to detect temporal shifts in user interests or behavioural dynamics without dedicated architectural or training interventions. The findings highlight the need for an enhanced agent design that explicitly accounts for temporal information and models user behaviour as a dynamic and sequential process.

To address the challenges, we propose DyTA4Rec, a novel LLM-based simulator for recommendation systems that models both static and dynamic features and captures temporal behaviour patterns for informed decision-making. Our main contributions are:
\begin{itemize}[leftmargin=0.5cm]
    \item DyTA4Rec presents a temporal-aware simulation framework that improves the behavioural fidelity and temporal consistency of agent interactions through sequential reasoning.
    \item The framework integrates a dynamic profile updater to capture real-time short-term user modelling and a self-adaptive aggregator for combining multi-dimensional behavioural signals.
    \item Experiments on the MovieLens-1M dataset show that DyTA4Rec aligns agent behaviour with real user patterns in terms of accuracy and believability at both group and individual levels.
\end{itemize}

\begin{figure*}[t]
    \centering
    \includegraphics[width=0.90\linewidth]{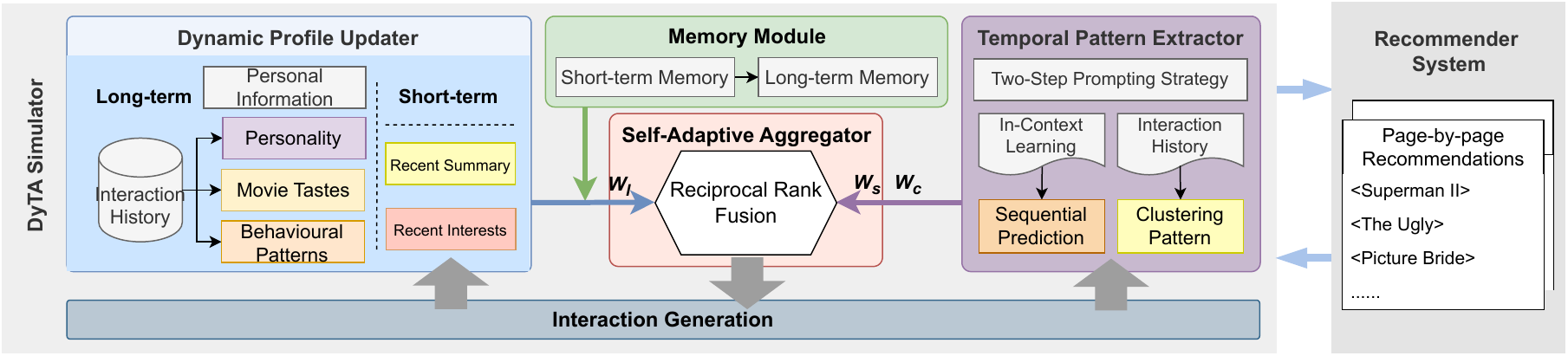}
    \caption{An overview of the proposed DyTA4Rec, comprising an LLM-based generative agent and a recommender system. The recommender system is designed as a flexible component that can be reconfigured to support different objectives. During simulation, the agent replicates real user behaviour, enabling realistic interactions with the recommender system.}
    \label{fig:architecture}
\end{figure*}

\section{DyTA4Rec}
The architecture of DyTA4Rec is shown in Fig.~\ref{fig:architecture}. It comprises four main modules: (1) dynamic profile updater, (2) temporal pattern extractor, (3) self-adaptive aggregator, and (4) memory module. 
The following subsections provide a detailed explanation.

\subsection{Dynamic Profile Updater}

The user profile integrates both static, long-term attributes (e.g., demographic information, preferences, and behavioural patterns) and dynamic, short-term features that reflect recent interactions and evolving interests. Long-term preferences typically represent stable user interests, whereas short-term preferences are more specific and context-dependent, capturing temporary intent~\cite{sun2020go}. This hybrid representation enables the agent to form a comprehensive and temporal-aware understanding of user behaviour, thereby supporting more accurate and personalised simulation.

Long-term features are initialised based on the MovieLens-1M (ML-1M) dataset \cite{harper2015movielens}. 
% We preserve user metadata, including ID, gender, and postcode. 
% While we use ML-1M as an illustrative example, the framework is generalisable and can be applied to other datasets by adapting to the available user attributes and interaction records. 
To enrich user profiles beyond basic metadata, we employ LLMs to extract personalities and item-level preferences from each user's historical interactions. In addition, we compute statistical patterns to capture rating tendencies (e.g., distributions over popular or highly rated items) and viewing behaviours (e.g., preferences for popular or highly rated content). These patterns guide agent's decision-making during simulation. The modular design of the dynamic updater enables flexible incorporation of dataset-specific features, enhancing its adaptability to a wide range of application scenarios.
Specifically, we leverage LLMs to extract and summarise users' recent behaviours and evolving interests from their most recent interaction history, serving as short-term features updated every $n$ rounds.

% Seems not complete. Could add any equations?
% Is there any room to put the prompt? Or put it in the provided code link.

\subsection{Temporal Pattern Extractor}
We propose a Temporal Pattern Extractor (TPE) to capture temporal dynamics and sequential patterns in user behaviour by analysing most recent interaction histories.
% short-term behavioural signals related to user preferences.

The process of TPE can be formulated as:
\begin{equation}
    \text{TPE}(H, C) = f^{act.}(f^{cluster.}(H)) + f^{seq.}(\text{ICL}(H))
\end{equation}
where $f$ denotes utilising LLMs for clustering, prediction, and decision-making. The function $f^{act.}$ represents the action mechanism of the agent, which may involve ranking candidates, selecting the next interaction, or performing other task-specific operations.

For each agent, we define the most recent $n$ interactions as $H = \{h_1, h_2, ..., h_n\}$, where each $h_t = (i_t, r_t, f_t)$ represents an interaction involving item $i_t$, a rating $r_t$, and a generated feeling $f_t$. During simulation, the recommender system presents content to the agent in a page-by-page manner, with each page containing a candidate set $C = \{c_1, c_2, ..., c_m\}$. Temporal information of $H$ is obtained from clustering $f^{cluster.}$ and sequential prediction $f^{seq.}$. $f^{cluster.}$ identifies recurring behaviour patterns and detects potential shifts in user interests. $f^{seq.}$ captures ordered dependencies in $H$ to infer the most likely next interaction. This combination enables agents to make time-consistent decisions that align with recent interaction trajectories.

Additionally, we adopt a two-step prompting approach for clustering pattern extraction to more effectively leverage the reasoning capabilities of LLMs. Specifically, the LLM performs an independent analysis, identifying key patterns or insights. Then, this intermediate analysis is provided as contextual information for the LLM to execute the specific downstream task, such as prediction or ranking. This structured process enhances the quality of reasoning and improves task performance by separating understanding from decision-making (as shown in Figure~\ref{fig:clustering_strategies}).

We leverage LLMs through in-context learning (ICL) to enable such temporal reasoning without additional training. LLMs have demonstrated the ability to adapt to new tasks by conditioning on a few prompt-based examples, eliminating the need for model retraining~\cite{wei2021finetuned,dong2022survey}. In our framework, both temporal clustering and sequential prediction functions are instantiated directly from the interaction history using LLMs.

\subsection{Self-Adaptive Aggregator}

% The strategy for aggregating multiple behavioural signals plays an important role in determining the quality of the final action decision. 
We design an aggregation approach self-adaptive aggregator (SAA) for integrating complementary behavioural information (i.e., long- and short-term user traits and interests, sequential patterns, and temporal clustering patterns and balancing their relative importance. We implement and compare Borda Count (BC) and Reciprocal Rank Fusion (RRF) \cite{cormack2009reciprocal} as aggregation methods. Compared to averaging or majority voting with fixed weights, our method assigns personalised weights based on whether the interaction history contains consistent temporal or sequential patterns. This enables more context-sensitive information fusion and decision generation.

User profile and memory information serve as the foundation of decision-making, hence the corresponding weight $W_l$ is set to 1. For temporal reasoning, LLMs are employed to determine whether the interaction history exhibits clear sequential patterns or temporal clusters. A weight $W_s$ is allocated to the sequential prediction output when sequential dependencies are observed, while a weight $W_c$ is applied to the clustering-based outcome in the presence of discernible temporal clustering patterns.
% A weight $W_s$ is assigned to the sequential prediction output when sequential dependencies are identified. A weight $W_c$ is assigned to the clustering-based result when temporal clustering patterns are detected.

The aggregation process is defined as:
\begin{equation}
    \text{Aggregator}(R) = W_l \times r_l + W_s \times r_s + W_c \times r_c
\end{equation}
where $R = \{r_l, r_s, r_c\}$ denotes the results generated by different modules. The term $r_l = w_i \times r^1_l$ represents the output of the user profile and memory module, $r_s$ represents the output of sequential prediction, and $r_c$ represents the output of clustering analysis.

\subsection{Memory Module}
Memory constitutes the core of the human decision-making process \cite{wang-etal-2024-recmind, zhang2024survey}. Inspired by previous work \cite{zhang2024generative, wang2025user}, we design a specialised memory module for LLM-based simulation in recommendation. Our memory module consists of two components: short-term memory and long-term memory. Specifically, the short-term memory captures the recent interaction history directly and tracks the emotional feelings, facilitating the system to adapt to the dynamic environment. In contrast, the long-term memory stores high-level and informative patterns that remain consistent over time, which are extracted and summarised from short-term memory by LLM.

\section{Experiments}
% In this section, to fully demonstrate the effectiveness of our proposed simulation framework, we conduct a set of extensive experiments to study the following research questions:
% (1) To what extent is it possible to leverage LLM to track user's short-term features dynamically based on the recent interaction history?
% (2) Can LLM-empowered temporal reasoning improve the alignment between agents behaviour and real human behaviour?
% (3) Is ensemble-based prompting method more effective than single prompting method?

\subsection{Experimental Setup}

We conduct experiments on the benchmark recommendation dataset ML-1M~\cite{harper2015movielens}. We adopt the GPT-4o-mini, due to budget constraints, for simulations by default. To reduce output variability from the language model, the temperature is set to 0.1 and the top-$p$ value is set to 0.9. All reported results are averaged over three independent runs to ensure reliability and consistency.

We consider four baseline models:  
(1) Random, a random ranking strategy;  
(2) RecAgent~\cite{wang2025user}, an LLM-based agent simulation framework;  
(3) BM25~\cite{robertson2009probabilistic}, a text-based retrieval algorithm that ranks items based on their textual similarity to user preferences; and  
(4) UniSRec~\cite{hou2022towards}, a pre-trained sequential recommendation model.

The evaluation metrics include the normalised discounted cumulative gain ({nDCG@K}), where $K \in \{5, 10\}$, and the top-N hit rate ({HR@3}). Following previous studies \cite{zhou2020s3,hou2022towards,zhang2024agentcf}, we adopt the leave-one-out evaluation strategy, where the last item in each historical interaction sequence is treated as the ground-truth item.

We construct ranking candidates by including one ground-truth item and nine negative samples. Empirical research~\cite{collins2018position,hou2024large} identifies position bias in RSs, where users are more likely to select items near the top of the recommendation list. To mitigate this bias, we apply a simple Direct Prompting strategy to explicitly instruct that candidate positions carry no significance.
% and (2) Bootstrapping, which randomly shuffles the candidate list, performs ranking over three rounds, and aggregates the results across rounds~\cite{hou2024large}.

\begin{table}[t]
  \centering
    \caption{Performance comparison on the ML-1M dataset. The first block reports baseline results. A tailored version of RecAgent is evaluated on ML-1M since the social relationship information of dataset is not provided. The second block presents ablation studies of DyTA4Rec by varying behavioural inputs and removing the SAA. The third block shows the performance of the DyTA4Rec using different aggregation strategies. The best result in each row is highlighted in \textbf{bold}, and the second-best result is \underline{underlined}.}
  \label{tab:ranking_evaluation}
    \small{
  \begin{tabular}{l|ccc}
    \toprule
    Model & nDCG@5 & nDCG@10  & HR@3 \\
    \midrule
    Random             & 0.326 & 0.509 & 0.240 \\
    BM25  & 0.260 & 0.428 & 0.242  \\
    UniSRec        & 0.340 & 0.484 & 0.365 \\
    RecAgent & 0.431 & 0.571 & 0.391 \\
    \midrule
    Long-term           & 0.434 & 0.568 & 0.360 \\
    Long- \& Short-term & 0.467 & 0.599 & 0.400 \\
    Sequential         & 0.466 & 0.584 & 0.418 \\
    Clustering         & 0.462 & 0.593 & 0.409 \\
    Sequential+Long-term & 0.494 & 0.622 & 0.458 \\
    DyTA4Rec w/o SAA (BC) & 0.489 & 0.602 & 0.427 \\
    DyTA4Rec w/o SAA (RRF)& 0.457 & 0.585 & 0.396 \\
    \midrule
    DyTA4Rec (BC)  &\underline{0.514} & \underline{0.632}  & \underline{0.467} \\
    DyTA4Rec (RRF)  & \textbf{0.551} & \textbf{0.639}  & \textbf{0.480} \\
    \bottomrule
  \end{tabular}}
\end{table}

\begin{figure*}[t]
	\centering
        \begin{minipage}[c]{0.195\linewidth}
		\centering
        \includegraphics[width=\linewidth]{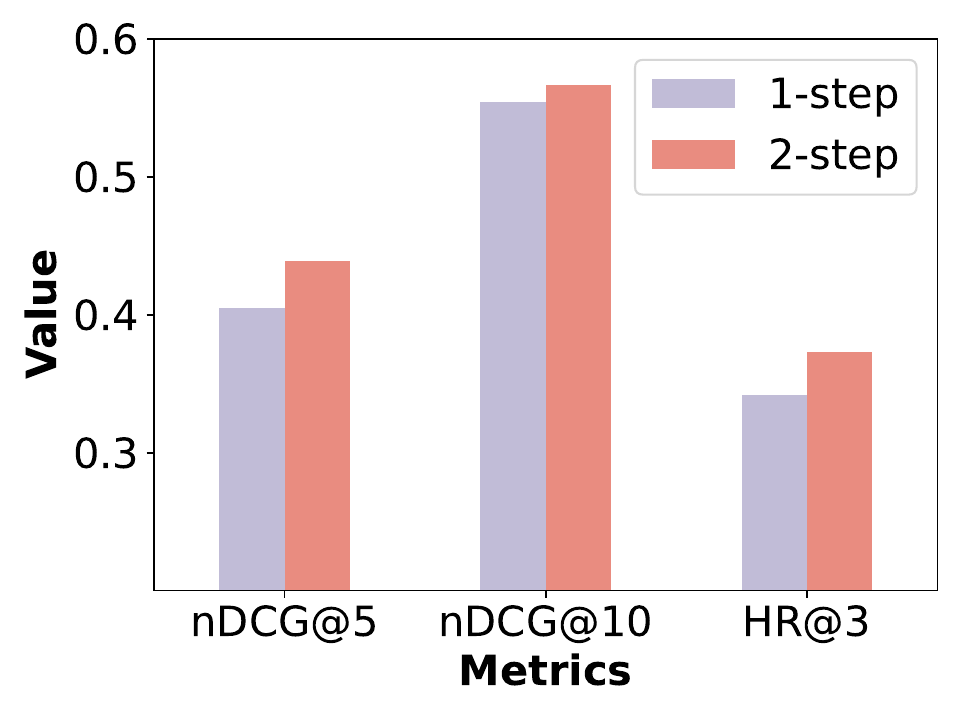}
		\subcaption{}
		\label{fig:clustering_strategies}
	\end{minipage}
        \begin{minipage}[c]{0.195\linewidth}
		\centering
		\includegraphics[width=\linewidth]{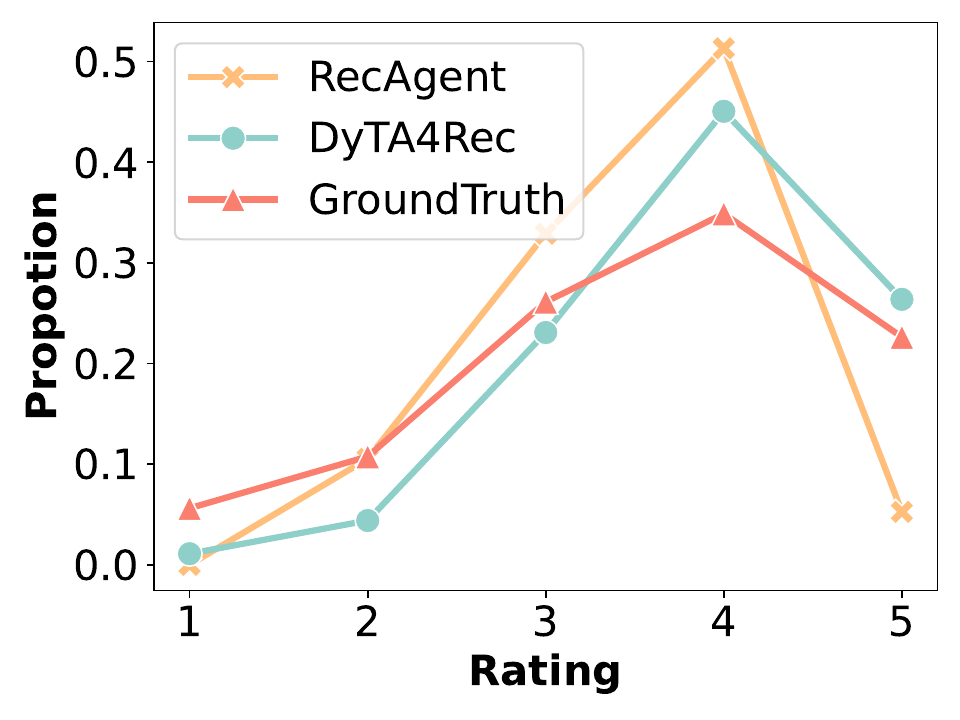}
		\subcaption{}
		\label{fig:rating_distribution}
	\end{minipage}
	\begin{minipage}[c]{0.195\linewidth}
		\centering
		\includegraphics[width=\linewidth]{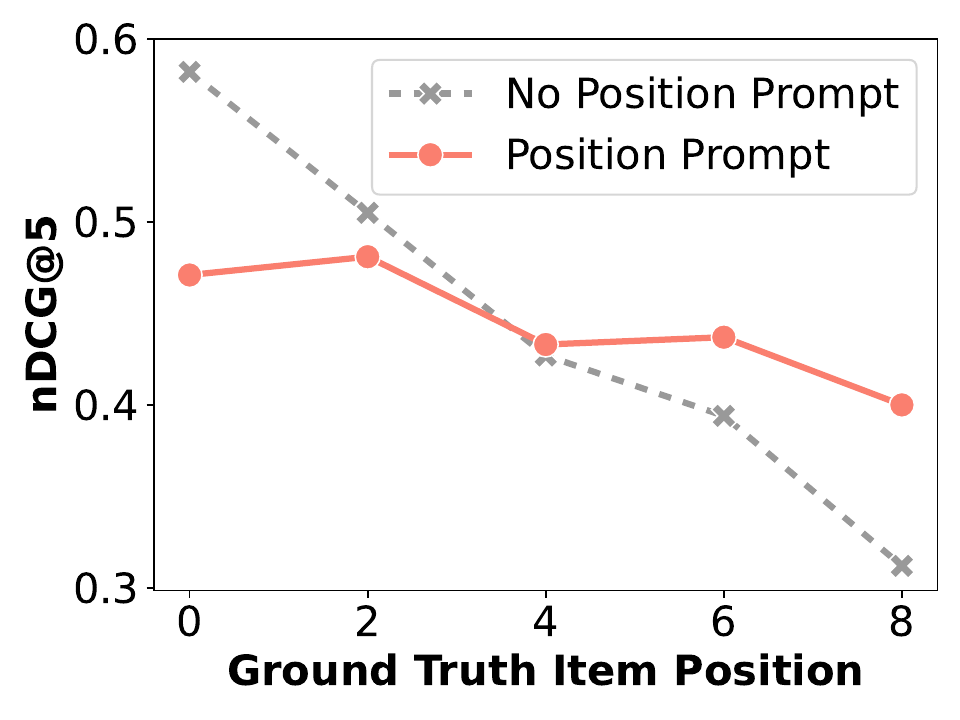}
		\subcaption{}
		\label{fig:position_bias}
	\end{minipage}
	\begin{minipage}[c]{0.195\linewidth}
		\centering
		\includegraphics[width=\linewidth]{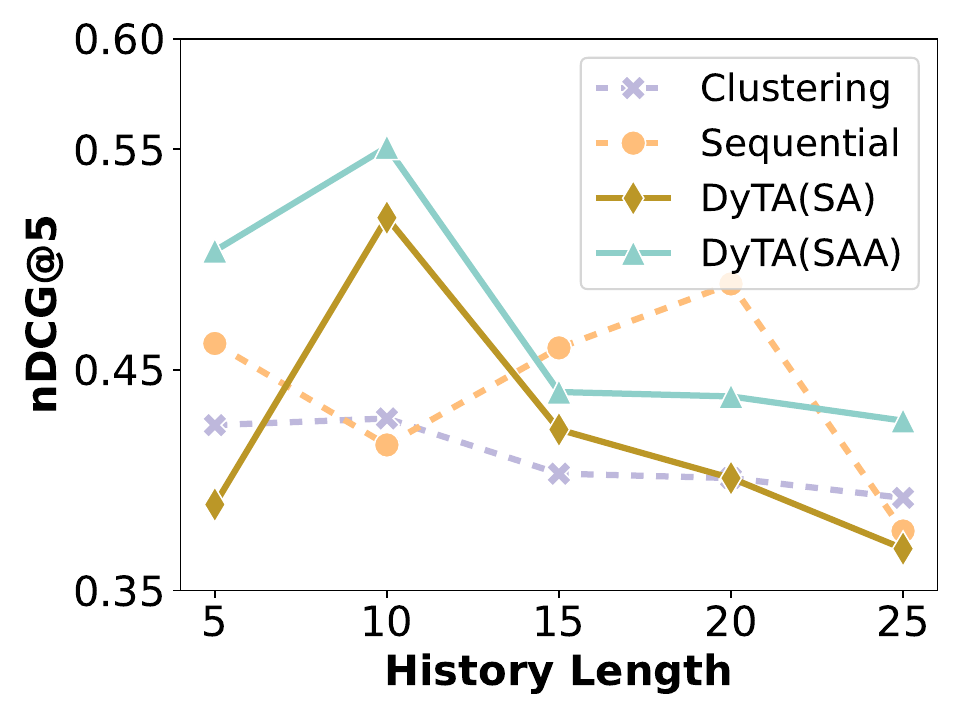}
		\subcaption{}
		\label{fig:his_length}
	\end{minipage}
	\begin{minipage}[c]{0.195\linewidth}
		\centering
		\includegraphics[width=\linewidth]{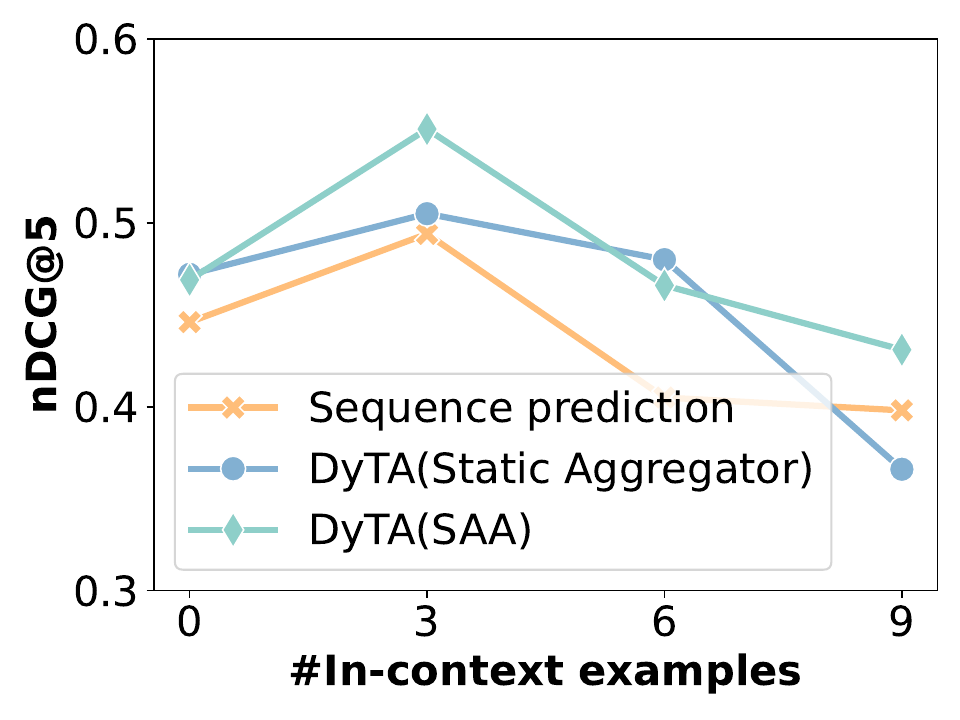}
		\subcaption{}
		\label{fig:ICL_number}
	\end{minipage} 
 %        \begin{minipage}[c]{0.49\linewidth}
	% 	\centering
	% 	\includegraphics[width=\linewidth]{figures/voting_strategies.pdf}
	% 	\subcaption{Voting strategies comparison.}
	% 	\label{fig:voting_strategies}
	% \end{minipage}
    \caption{
    Performance comparisons across various settings: 
    (a) ranking performance under single-step and two-step prompting strategies; 
    (b) rating distribution alignment between ground truth and simulated data (RecAgent vs. DyTA4Rec); 
    (c) impact of ground-truth item position on simulation outcomes; 
    (d) effect of interaction history length; 
    (e) effect of ICL example count.
    }
\end{figure*}
\subsection{Main Results} 
We assess the \textbf{ranking ability} of each agent. Main results are presented in Table~\ref{tab:ranking_evaluation}. DyTA4Rec consistently outperforms the baseline models in next-item prediction. The traditional retrieval-based method BM25 performs poorly across all metrics, reflecting the limitations in capturing user intents. UniSRec performs better than BM25 due to its temporal modelling ability. RecAgent shows further improvement by incorporating generative reasoning. Notably, DyTA4Rec with RRF achieves the best results, with 0.551 in nDCG@5, 0.639 in nDCG@10, and 0.480 in HR@3, demonstrating the effectiveness of our temporal-aware and adaptive aggregation design.

Beyond ranking performance, we further evaluate the \textbf{behavioural alignment} between simulated agents and real users at a macro level. Following the setup in~\cite{zhang2024generative}, we assess rating distribution consistency on ML-1M. Figure~\ref{fig:rating_distribution} compares the ground-truth rating distribution with those generated by the baseline simulator~\cite{wang2025user} and our proposed simulator. 
% Compared to the baseline model, the rating distribution produced by DyTA4Rec shows closer alignment with the ground truth, highlighting stronger consistency between simulated agent behaviour and real human preferences at the group level.
DyTA4Rec outperforms the baseline models in aligning with actual ratings, showing stronger consistency between simulated agents and real human behaviour at the group level.

Our study also investigate the impact of position bias within simulations. Figure~\ref{fig:position_bias} shows that the performance of LLM-based agents is highly sensitive to the recommended items position, as the selection is easily influenced by items order. Without explicit prompts indicating that item positions are random or irrelevant, agents often overvalue top-ranked options. This reflects real-world position bias where top-listed items receive disproportionate attention and engagement~\cite{chen2023bias}. 
To address this, we design an explicit prompt to inform the agent of the randomness of item positions, effectively reducing position bias in its decisions.
% To mitigate this issue, we adopt an explicit prompting strategy to inform the agent of the randomness of item positions. This simple strategy effectively reduces the impact of position bias on agent decisions.

\subsection{Ablation Study}
We conduct an ablation study to examine the contribution of different components in DyTA4Rec (second block, Table~\ref{tab:ranking_evaluation}). The user profile integrates both long-term and short-term features. Long-term features are initialised from real-world data and remain fixed throughout the simulation, while short-term features are dynamically generated and updated by the profile updater based on recent interactions. To assess the effect of dynamic profiling, we compare two configurations: using only long-term features ({Long-term}) and combining long-term with short-term features ({Long \& Short-term}). 

We further study the impact of each temporal reasoning strategy. The {Sequential} and {Clustering} variants enable only the sequential prediction or temporal clustering module, respectively, while {Sequential + Long-term} combines sequential reasoning with long-term profiling. In addition, we disable the SAA and apply static aggregation (SA) strategies, namely BC and RRF, to examine the importance of adaptive weighting.

Table~\ref{tab:ranking_evaluation} shows that incorporating short-term features yields clear improvements over using long-term profiling alone (e.g., nDCG@5 increases from 0.434 to 0.467), confirming the value of dynamic features tracking. Similarly, enabling sequential or clustering reasoning individually improves performance, and combining sequential signals with long-term profiling leads to further gains. Removing the SAA also results in performance degradation, demonstrating the effectiveness of adaptive strategy integration. These results highlight the importance of both temporal reasoning and dynamic profiling in achieving high-quality simulation.

% \begin{figure}[t]
% 	\centering
% 	\begin{minipage}[c]{0.49\linewidth}
% 		\centering
% 		\includegraphics[width=\linewidth]{figures/rating_distribution.pdf}
% 		\subcaption{}
% 		\label{fig:rating_distribution}
% 	\end{minipage}
% 	\begin{minipage}[c]{0.49\linewidth}
% 		\centering
% 		\includegraphics[width=\linewidth]{figures/position_bias.pdf}
% 		\subcaption{}
% 		\label{fig:position_bias}
% 	\end{minipage}
%     \caption{(a) Comparison of rating distributions between ground truth and simulated data generated by RecAgent and DyTA4Rec. 
%     (b) Impact of ground-truth item position within the recommendation list on simulation outcomes.}
% \end{figure}

\subsection{Parameter Analysis}

We investigate the impact of two key parameters in DyTA4Rec: the length of the interaction history and the number of in-context examples used for sequential behaviour modelling. These parameters influence how effectively the agent can capture temporal dependencies and recent user intents during the simulation.

First, we examine the impact of interaction history length on performance. As shown in Figure~\ref{fig:ICL_number}, the performance of all three agents tends to decline as the length of the interaction history increases. This may be attributed to two aspects: (1) limited ability of LLMs to extract and reason over long behavioural sequences although they can process long textual inputs~\cite{liu2023chatgptgoodrecommenderpreliminary, hou2024large}; and (2) long sequence may distract LLMs' attention from most recent interactions, which are often more informative for decision-making and more accurately reflect the user’s current intent.
% focusing on the most recent interactions is often more informative for decision-making, as they better reflect the user’s current intent.

Second, we study the influence of ICL by varying the number of in-context examples in the prompt. The interaction history length is set to 10 and the most recent $k$ items are selected as ICL examples, with $k \in \{0, 3, 6, 9\}$. As shown in Figure~\ref{fig:ICL_number}, using 3 examples achieves the best overall performance. This suggests that a moderate number of examples is sufficient to enhance temporal reasoning, while excessive prompts may introduce noises.

\section{Conclusion and Future Work}

In this paper, we present DyTA4Rec, a novel LLM-based simulator that enhances the temporal awareness of agents in recommendation scenarios. By integrating a dynamic profile updater, a temporal pattern extractor, and a self-adaptive aggregator, DyTA4Rec captures both static and evolving user behaviour patterns for more realistic simulations. Experiments demonstrate improved alignment with real user preferences at both individual and group levels. However, the effectiveness of LLM-based agents remains sensitive to prompt design and prone to overconfident outputs.
Future work will explore integrating an extra model to guide the output generation, aiming to mitigate the prompt sensitivity. 
% Another potential direction for improvement involves integrating Retrieval-Augmented Generation (RAG) techniques to enhance grounding, reduce hallucinations, and enable deeper item-level interactions.
We also plan to generalise our simulation framework on different benchmark datasets across diverse domains to assess its generalisability and robustness.

\begin{acks}
This research was conducted by the ARC Centre of Excellence for Automated Decision-Making and Society (ADM+S, CE200100005), and funded fully by the Australian Government through the Australian Research Council.
\end{acks}

% \newpage
\section*{GenAI Usage Disclosure}

% \chenglong{Please review and update this section before submission.}

This research complies with the CIKM 2025 GenAI usage policy. The authors disclose the following use of Generative AI (GenAI) tools during the research process:

\begin{itemize}
  \item \textbf{Writing:} ChatGPT (GPT-4, OpenAI) was used to assist in proofreading, rephrasing technical sentences, and improving the clarity of the manuscript. All substantive content, including ideas, methods, results, and analysis, was written and verified by the authors.
  \item \textbf{Code:} No GenAI tools were used to generate any code in this research. All code was developed by the authors.
  \item \textbf{Data:} No GenAI tools were used to generate or augment the data used in this research. All datasets were obtained from publicly available sources as described in the paper.
  \item \textbf{Experiments and Analysis:} No GenAI tools were used to generate experimental results or statistical analyses.
\end{itemize}

The authors confirm that all intellectual contributions are original and that the use of GenAI tools did not compromise the scientific integrity or originality of the work.
\bibliographystyle{ACM-Reference-Format}
\balance
\bibliography{main}

% \appendix

\end{document}